\documentclass[10pt,conference]{IEEEtran}

\usepackage{amsmath, amssymb, bm, cite, epsfig, psfrag}
\usepackage{graphicx}
\usepackage{subcaption}
\usepackage[top    = 0.7in,
bottom = 0.95in,
left   = 0.7in,
right  = 0.7in]{geometry}
\usepackage{array}
\usepackage{textcomp} 
\usepackage{longtable}
\usepackage{supertabular,booktabs}
\usepackage{authblk}
\usepackage{multirow}
\usepackage[usenames,dvipsnames]{xcolor}
\usepackage{etoolbox}
\usepackage{pbox}
\usepackage{romannum}
\usepackage{enumerate}
\usepackage{colortbl}
\usepackage{url}
\usepackage[switch]{lineno} 
\usepackage{bm}
\usepackage{float}
\usepackage{fixltx2e}
\usepackage{fancyhdr}

\graphicspath{{figures/}}
\fancypagestyle{firststyle}{
	\fancyhf{}
	\fancyhead[L]{O. Kanhere, A. Chopra, A. Thornburg, T. S. Rappaport, and S. S. Ghassemzadeh ``Performance Impact Analysis of Beam Switching in Millimeter Wave Vehicular Communications,''  2021 \textit{IEEE 93rd Vehicular Technology Conference (VTC-Spring)}, April 2021, pp. 1-7.
	}                           

}
\IEEEoverridecommandlockouts

\begin{document}
\title{Performance Impact Analysis of Beam Switching in Millimeter Wave Vehicular Communications}	

\author[1]{Ojas Kanhere 	\thanks{This research is supported by the NYU WIRELESS Industrial Affiliates Program, AT\&T Labs, and National Science Foundation (NSF) Research Grants: 1909206.}}
\author[2]{Aditya Chopra}
\author[2]{Andrew Thornburg}
\author[1]{Theodore S. Rappaport}
\author[2]{\\Saeed S. Ghassemzadeh}
\affil[1]{NYU WIRELESS, NYU Tandon School of Engineering, \{ojask, tsr\}@nyu.edu}
\affil[2]{AT\&T Labs, \{aditya\_chopra, andrew\_thornburg, 
		saeed\}@labs.att.com}

\maketitle
	\thispagestyle{firststyle}
\begin{abstract}
Millimeter wave wireless spectrum deployments will allow vehicular communications to share high data rate vehicular sensor data in real-time. The highly directional nature of wireless links in millimeter spectral bands will require continuous channel measurements to ensure the transmitter (TX) and receiver (RX) beams are aligned to provide the best channel. Using real-world vehicular mmWave measurement data at 28 GHz, we determine the optimal beam sweeping period, i.e. the frequency of the channel measurements, to align the RX beams to the best channel directions for maximizing the vehicle-to-infrastructure (V2I) throughput. We show that in a realistic vehicular traffic environment in Austin, TX, for a vehicle traveling at an average speed of 10.5 mph, a beam sweeping period of 300 ms in future V2I communication standards would maximize the V2I throughput, using a system of four RX phased arrays that scanned the channel 360 degrees in the azimuth and 30 degrees above and below the boresight. We also investigate the impact of the number of active RX chains controlling the steerable phased arrays on  V2I throughput. Reducing the number of RX chains controlling the phased arrays helps reduce the cost of the vehicular mmWave hardware while multiple RX chains, although more expensive, provide more robustness to beam direction changes at the vehicle, allowing near maximum throughput over a wide range of beam sweep periods. We show that the overhead of utilizing one RX chain instead of four leads to a 10\% drop in mean V2I throughput over six non-line-of-sight runs in real traffic conditions, with each run being 10 to 20 seconds long over a distance of 40 to 90 meters.

\end{abstract}
    
\begin{IEEEkeywords}
mmWave; beam management; channel sounding; phased arrays; V2X; V2V; 5G; sidelink
\end{IEEEkeywords}

\section{Introduction}\label{sec:Introduction}

Modern automated vehicles require hundreds of sensors to ensure road safety \cite{Choi_2016}. Light detection and ranging (LIDAR) sensors generate a 3-D point cloud of the environment, allowing the vehicle to detect landmarks. The inertial measurement unit (IMU) measures the linear acceleration and rotation of the vehicle, which when combined with data from the global positioning system (GPS) receiver provides an estimate of the position and orientation of the vehicle. Radio frequency (RF) radars detect surrounding objects and are used for blind-spot detection \cite{Choi_2016}. 

By sharing the information gathered by a vehicle with surrounding vehicles and the infrastructure, cooperative applications such as platooning and smart traffic monitoring are possible \cite{Shimizu_2018}. Sharing sensor data between vehicles enables cooperative perception, wherein the field of view of the vehicle is expanded by sensor data from nearby vehicles \cite{Shimizu_2018}. Cooperative sensing reduces accidents by allowing vehicles to use awareness from surrounding vehicles, ensuring autonomous vehicles can safely perform traffic maneuvers. A 50\% reduction in crashes, injuries, and fatalities could be achieved by assisting drivers in making left turns and warning drivers about potential collisions at intersections \cite{USDoT}.

Today's spectrum allocations for vehicular communications are limited. For more than twenty years, 75 MHz in the 5.9 GHz band (5.85-5.925 GHz) was dedicated for intelligent transportation and vehicle safety systems using the Dedicated Short Range Communications (DSRC), an IEEE 802.11p-based wireless communication technology which supports a maximum bit rate of 27 Mb/s \cite{Perfecto_2017}. In 2020, the United States Federal Communications Commission reallocated a majority of this bandwidth away from connected vehicle technologies, leaving only 30 MHz (5.850-5.895 GHz) for cellular vehicle-to-everything (C-V2X) applications \cite{fcc_cv2x}. To meet the gigabit-per-second data rate requirements for vehicular sensor data sharing such as 3D LIDAR environmental imaging and high definition video feeds, a spectral bandwidth of hundreds of MHz would be required, for which the millimeter wave (mmWave) band is a feasible solution \cite{Choi_2016}. The mmWave band supports channels with bandwidths several hundreds of MHz wide, thus enabling sensor data sharing required for advanced vehicle-to-everything (V2X) applications. IEEE 802.11bd, an evolution of 802.11p, is currently being developed, which could support high throughput communication in the unlicensed mmWave band at 60 GHz \cite{Naik_2019}. 

To compensate for the additional path loss in the first meter of propagation, mmWave systems employ high gain antenna arrays with narrow antenna beamwidth \cite{Rappaport_2013b,Rappaport_2015,Rappaport_2017}. Phased arrays at 28 GHz with antenna beamwidths as narrow as 7$^\circ$ may be utilized \cite{Chopra_2020}. As observed in \cite{MacCartney_2017}, narrow beamwidth antennas at mmWave are more likely to be blocked by surrounding vehicles, with a vehicular blockage loss of 10-20 dB measured at 28 GHz in \cite{Park_2017}. The receiver (RX) antenna must quickly adapt to changing channel conditions due to dynamic blockage caused by the increased relative velocity of the transmitter (TX) and RX in vehicular channels. In \cite{Rappaport_2017_b}, the authors show that an impeding blockage may be sensed by a small fluctuation in received signal due to the diffraction edge. Beam failure is said to occur when the  signal-to-noise ratio (SNR) of a beam falls below a predetermined threshold. The authors in \cite{Akoum_2018} suggested that beam failure recovery procedures must be triggered as soon as a single failed beam is detected, rather than wait until all the TX and RX beams fail. In \cite{Sun_2013}, the authors proposed combining the signal energy from multiple beams simultaneously in order to improve the signal-to-interference ratio. The mmWave RX must carefully select the beamforming direction and periodically update the selected direction based on measurements of the evolving vehicular channel. Various beam tracking algorithms have been suggested in prior work to ensure beam alignment and prevent signal outage. The authors of \cite{Guo_2019} introduced a machine learning based beam tracking approach, where the channel state information (CSI) was used to build a long short term memory (LSTM) prediction model. In \cite{Jayaprakasam_2017}, an extended Kalman filter (EKF) was used to estimate the complex channel gain in order to minimize the beam alignment error, assuming initial access provided accurate initial channel parameters. The authors of \cite{Fazliu_2020} proposed a heuristic beam management strategy, wherein the TX beams were pointed towards stationary vehicles at red traffic lights, in order to serve vehicles for a longer time. 

Along with the design of accurate beam tracking algorithms, knowledge of the rate at which the mmWave channel should be measured is required for efficient mmWave system design. The beam sweep period is the interval between two channel measurements in a particular beamforming direction. Work in \cite{Va_2017} developed the concept of beam coherence time, which is the duration over which the TX and RX beams stay aligned, after which beam realignment is required. The authors in \cite{Hur_2018} studied beam management by observing the effect of beam sweeping rate on received power in an outdoor measurement campaign at 28 GHz. The required beam switching rate can be reduced by long-term beamforming, wherein the RX is aligned with macro-level angular directions that vary at a much slower rate in comparison to individual multipath components \cite{Sun_2014}. Frequent channel measurements with a short beam sweep period ensure higher RX SNR since the beams remain well aligned, however, the beam switching time overhead potentially nullifies any gains from switching to the strongest beam. A base station (BS) beam management procedure was proposed in \cite{Wang_2020}, where regions far from the BS were covered by narrow beams, while wider beams were used to cover regions near the BS. Additionally, recognizing the sparsity of the mmWave channel, the beamforming codebook was pruned to a select few beam directions based on the geometry of the roads in order to reduce the overhead to 1.7\% \cite{Wang_2020}.

The remainder of this paper is organized as follows. Section \ref{sec:CS_design} describes the channel sounder design. A description of the vehicular channel measurement campaign is provided in Section \ref{sec:measurements}. Using real-world mmWave vehicular channel data as a basis, a system model is developed to determine the optimal beam sweep period at the vehicle to maximize vehicle-to-infrastructure (V2I) throughput in Section \ref{sec:rx_design}. Finally, conclusions and future work are given in Section \ref{sec:conclusion}.

%
\begin{figure}[]
	\centering
	\includegraphics[width=0.5\textwidth]{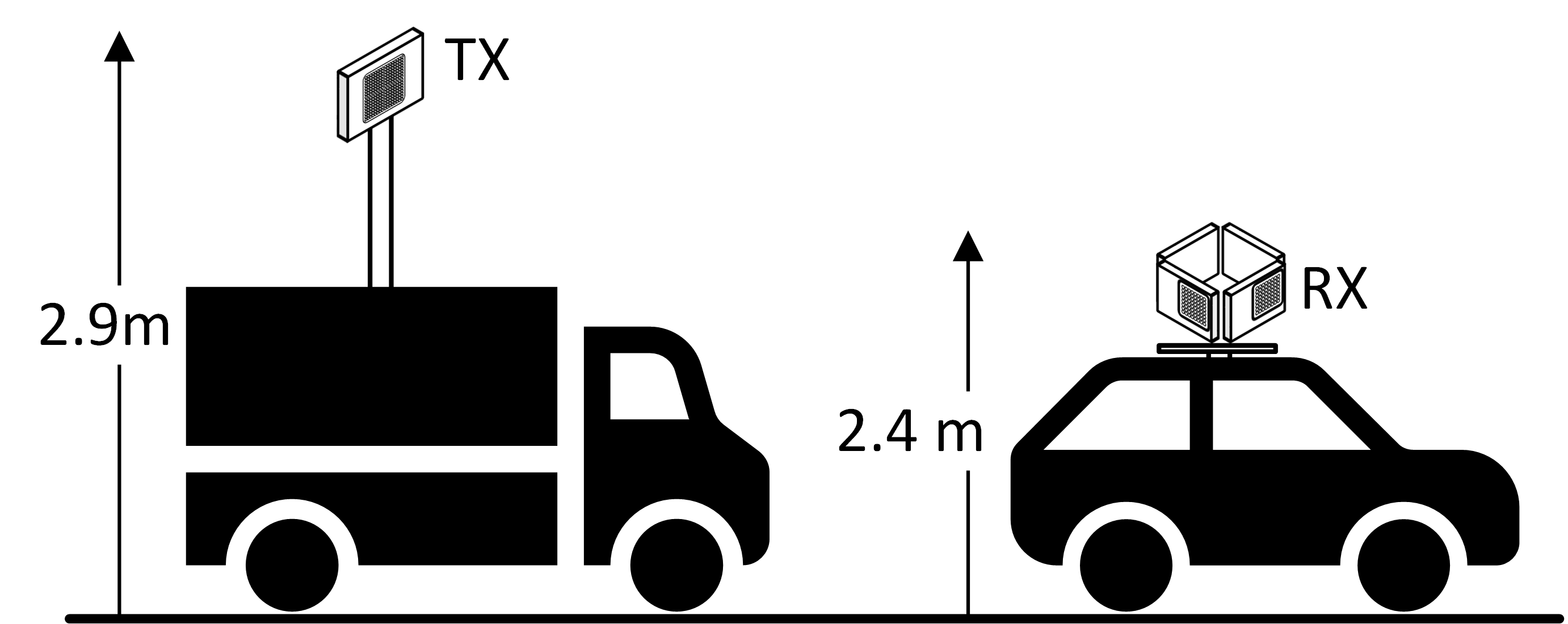}
	\caption{TX and RX vehicles used for channel sounding campaigns. Four RX antennas were arranged in a square pattern to provide 360$ ^\circ $ coverage, with a beam sweeping range of $ \pm$45$^\circ $. }
	\label{fig:vehicles} 	
\end{figure}

\section{Channel Sounder design}\label{sec:CS_design}
The novel Real-time Omni-Directional Channel Sounder (ROACH) developed at AT\&T Labs in Austin, is a wideband correlation type channel sounder \cite{Chopra_2020}. ROACH transmits a Zadoff-Chu (ZC) sequence\cite{hua_2014} of length 8192, sampled at 65.536 MHz, which results in a 125 $\mu$s long sounding sequence, equal to a 3GPP NR slot duration using a sub-carrier spacing of 120 kHz, in order to better relate channel measurements with NR system design parameters. At the TX, the ZC sequence is mixed with an intermediate frequency (IF) of 3.3 GHz and upconverted by 25 GHz, for an RF output frequency of 28.3 GHz. The RF signal is amplified and fed into the Anokiwave AWA-0134, a 256-element phased array module having a 3 dB beamwidth of 54.1$^\circ$ in the azimuth and a boresight gain of 36.8 dBi. Note that the boresight gain is the overall module gain, which is a combination of the gains from beamforming and power amplifiers on the module. A wide TX beamwidth was used to provide greater coverage without changing the TX pointing direction, however utilizing a narrower TX beamwidth may lead to different results since the smaller area illuminated by a narrower TX beamwidth will cause the vehicular channel will change more rapidly as the vehicle will move in and out of the smaller illuminated area more frequntly. In this paper we examine the effects of RX beamforming by switching only the beams of the ROACH RX that was mounted on a van at a height of 2.4 m as seen in Fig \ref{fig:vehicles}

\begin{figure}[]
	\centering
	\includegraphics[width=0.3\textwidth]{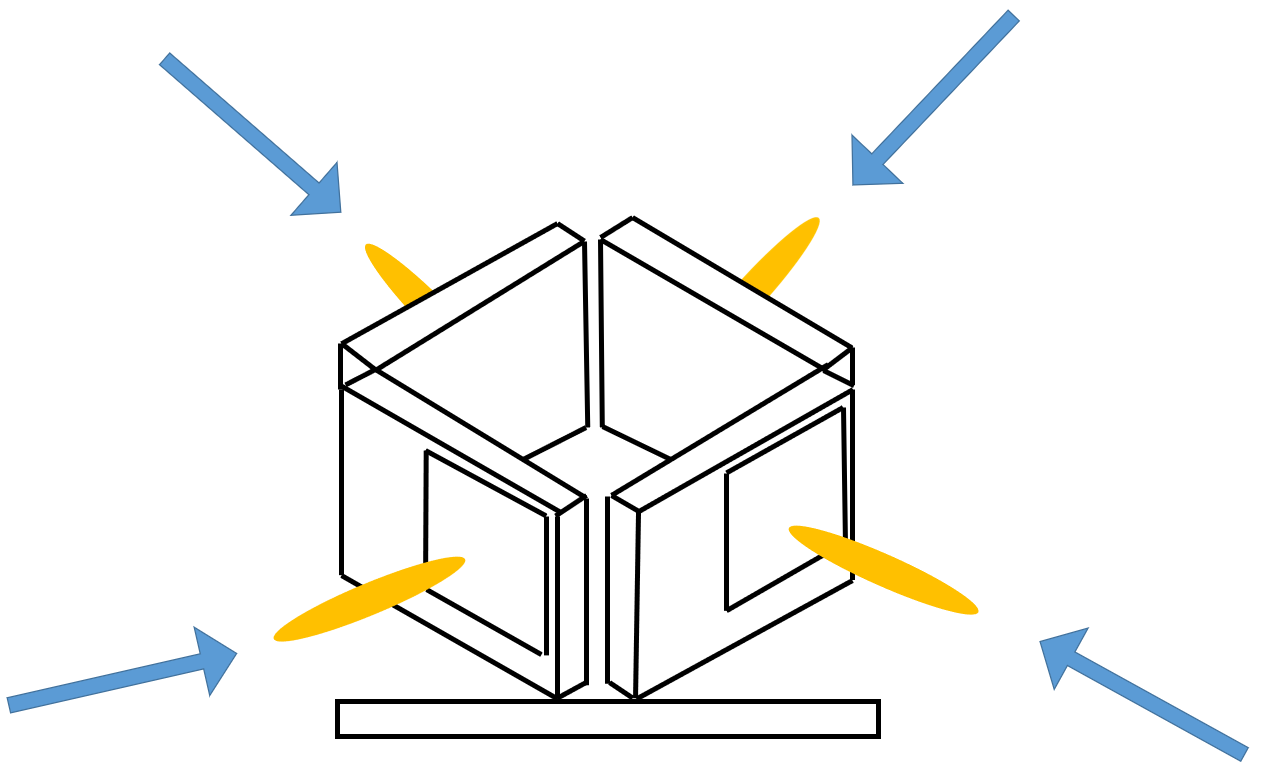}
	\caption{The four ROACH RX array faces jointly provide omnidirectional sensing capability \cite{Chopra_2020}. The 64-element phased arrays on each RX faces synchronously sweep through 50 predefined beamforming directions once every 6.25 ms.}
	\label{fig:RX_arrays} 	
\end{figure}

The RX used the Anokiwave AWMF-0129, a 64-element planar phased array module. The AWMF-0129 was operated with a 3 dB beamwidth of  16.8$^\circ $ and a boresight gain of 43.3 dBi. The boresight gain is the overall module gain, which is a combination of the gains from beamforming and low-noise amplifiers on the module. For omnidirectional reception in the azimuth, four Anokiwave AWMF-0129 arrays were placed in a square pattern as seen in Fig. \ref{fig:RX_arrays}, each covering 90$ ^\circ $ of the azimuthal plane. Consequently, the beam sweeping range for each array was chosen to be $\pm$45$^{\circ}$ in azimuth, and $\pm$30$^{\circ}$ in elevation from boresight. 

The received signals from each of the four arrays were fed to a four-input down-converter with a 25 GHz local oscillator producing four simultaneous IF signals. The received IF signals were cross-correlated with the ZC sequence to generate the power-delay profiles (PDPs) and received correlated power simultaneously on each of the four antenna array faces. The channel measurements were timestamped and sent to the host computer for immediate visualization and storage.

The four RX AWMF-0129s switched beams via serial-parallel interface (SPI) control. The 3.3 V transistor-transistor logic (TTL) output of a NI USRP-2955 general purpose input output (GPIO) subsystem was converted into four separate synchronized low voltage differential signal (LVDS) digital inputs, enabling synchronous beam switching for each of the four phased arrays. The 3-D spherical segment bounded by the $ \pm$30$^\circ $ elevation planes was tessellated by a 200 uniform hexagonal lattice with the beam directions at the centers of each hexagon (refer to \cite{Chopra_2020} for an illustration of the beam pattern). Each of the four phased arrays was assigned 50 beams, which were switched every sounding interval of 125 $\mu$s, resulting in an overall segment scan time of 6.25 ms. 

\begin{figure}[]
	\centering
	\includegraphics[width=0.5\textwidth]{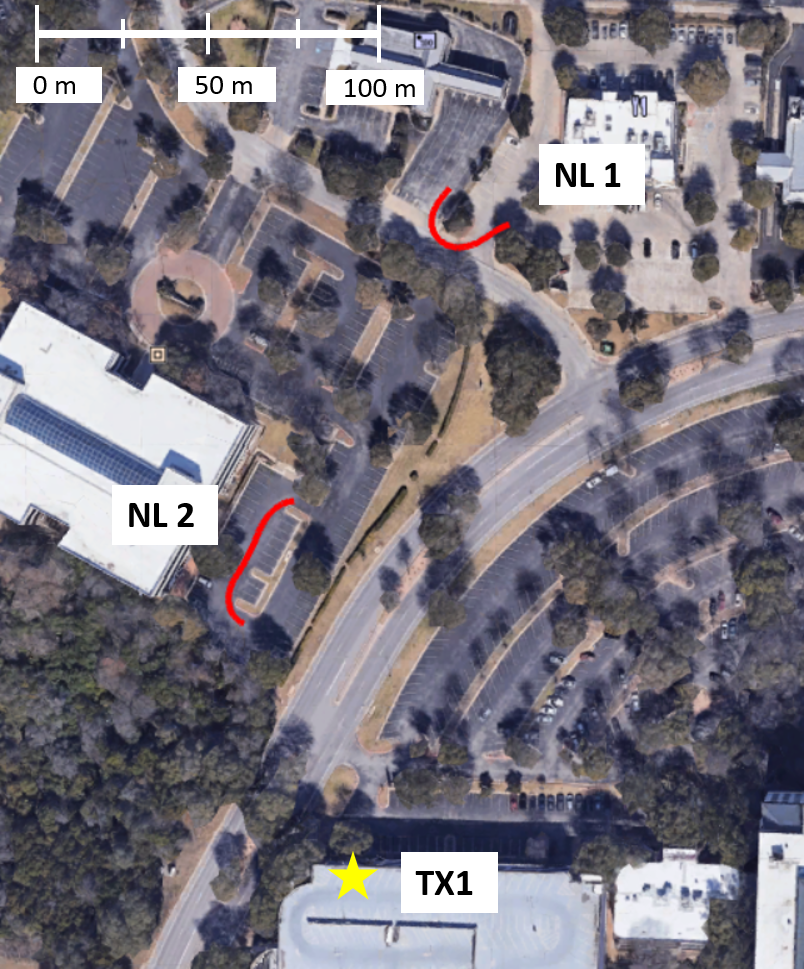}
	\caption{A map of the V2X Type I measurement environment, with the TX mounted on the roof of a building.}
	\label{fig:arboretum_scenarios} 	
\end{figure}
\begin{figure*}
	\centering
	\includegraphics[width=0.7\textwidth]{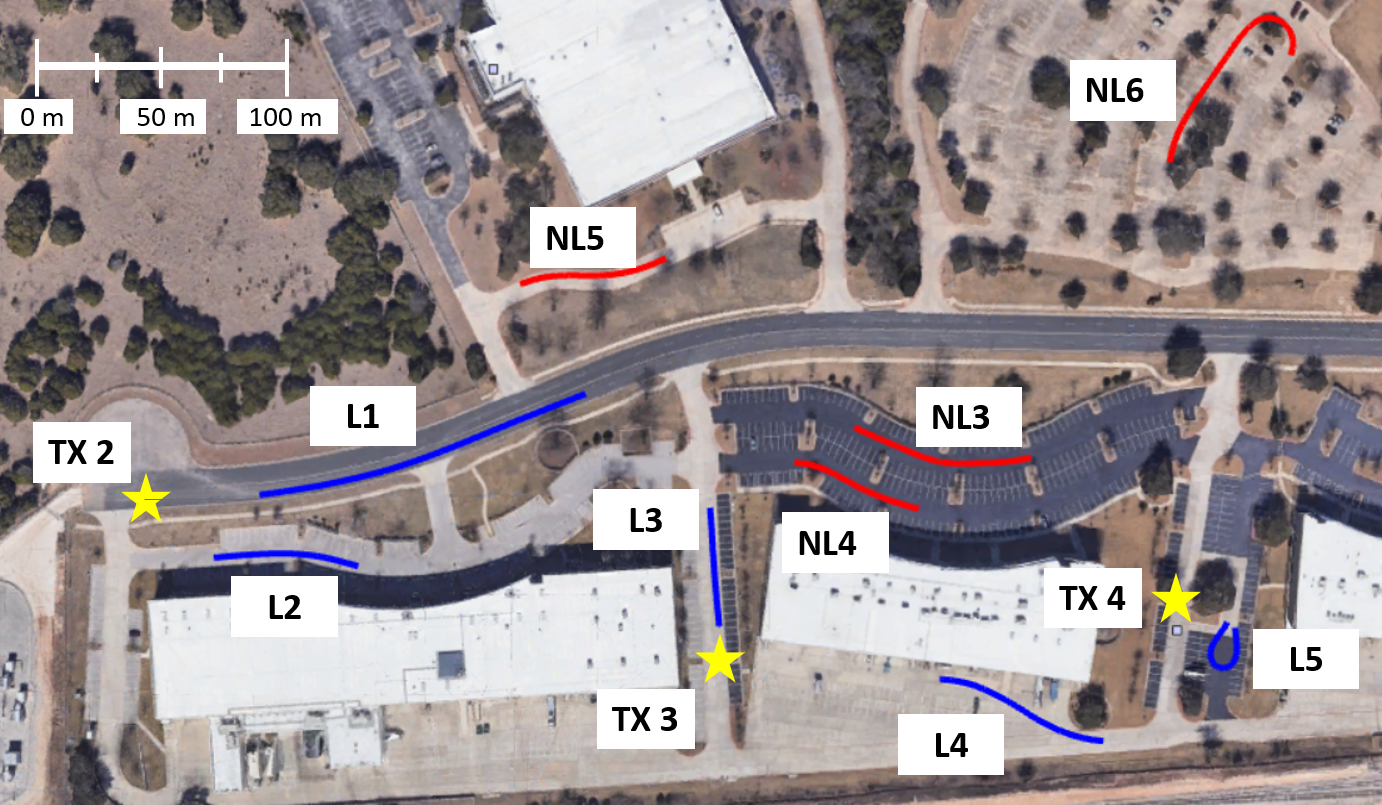}
	\caption{A map of the V2X Type II measurement environment, including a depiction of the scenarios of interest where the effect of beam sweeping period was studied. A list of the TXs serving each run are provided in Table \ref{tbl:Locations}}
	\label{fig:spectrum_scenarios} 	
\end{figure*}
\section{Channel Measurements}\label{sec:measurements}

The ROACH system was used to conduct V2I channel measurements in the 28 GHz frequency band. An experimental permit to temporarily transmit in this band was granted by the Federal Communications Commission (FCC)\cite{STA}.

 Unlike the vehicle-to-vehicle (V2V) scenario, where the TX and RX are at roughly the same height, the infrastructure TX is typically at a greater height. The V2I channel measurements were conducted in and around two outdoor urban light commercial areas containing offices and retail stores, cars, and light pedestrian traffic. The first area was adjacent to multiple shopping complexes and office buildings as depicted in Fig. \ref{fig:arboretum_scenarios}, while the second area contained office complexes and residential apartment buildings as seen in Fig. \ref{fig:spectrum_scenarios}. In the first area, the transmitter was mounted on an accessible roof of a building at a height of 15 m, while in the second area, the TX was mounted on an extendable mast on top of a stationary van at a height of 2.9 m (roughly at the height of light-posts) as shown in Fig. \ref{fig:vehicles}. The RX array system was mounted on the roof of a van, at a height of 2.4 m while the RX baseband system was mounted on a rack inside the van. 

\begin{figure}
	\centering
	\includegraphics[width=0.5\textwidth]{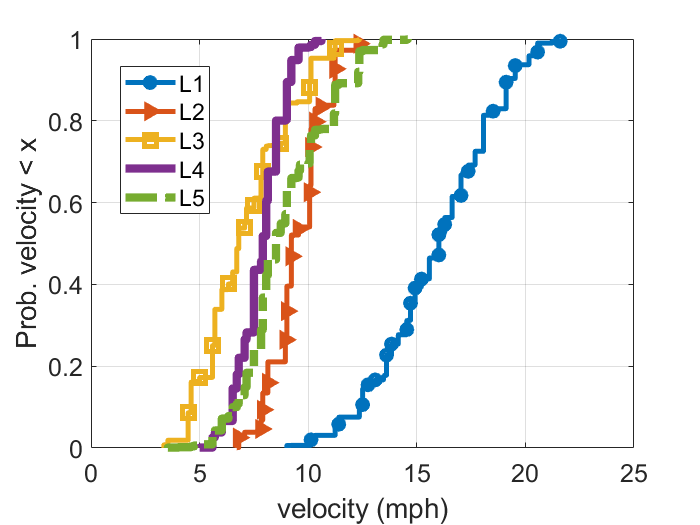}
	\caption{The velocity profile of the LOS scenarios. }
	\label{fig:LOS_velocity} 	
\end{figure}

\begin{figure}
	\centering
	\includegraphics[width=0.5\textwidth]{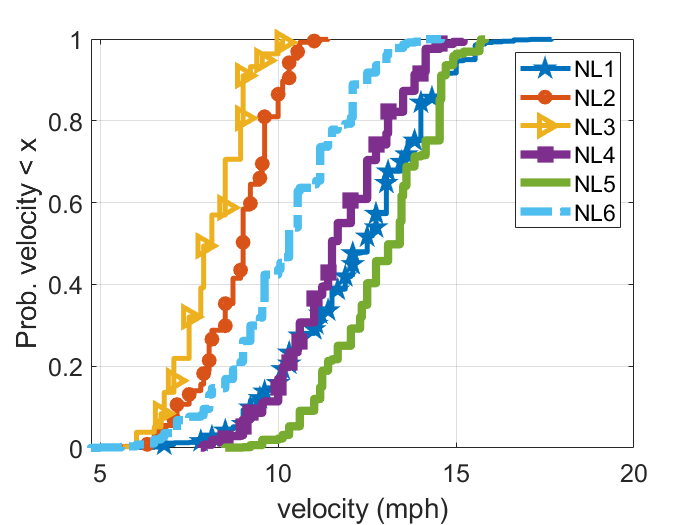}
	\caption{The velocity profile of the NLOS scenarios. }
	\label{fig:NLOS_velocity} 	
\end{figure}

The RX van was driven around on surface roads and shopping and office parking lots near the transmitter, ensuring that both line-of-sight (LOS) and non-line-of-sight (NLOS) locations were covered. Five LOS runs (L1, L2, L3, L4, and L5) and six NLOS runs (NL1, NL2, NL3, NL4, NL5, and NL6) were chosen for analysis in this paper, with each run 10-20 seconds long. The runs are marked on the maps in Fig. \ref{fig:arboretum_scenarios} and Fig. \ref{fig:spectrum_scenarios}, with the LOS runs marked in blue and the NLOS runs marked in red. TX locations are marked with yellow stars. Table \ref{tbl:Locations} lists the serving TX each run. In this paper only a subset of all measurements conducted in vehicular measurement campaign has been chosen for analysis to ensure that the vehicular channel was studied in dynamic vehicular channel conditions that would faithfully emulate future 5G V2I deployments, avoiding locations where the vehicle experienced complete signal outage and locations too close to the TX (with very high SNR), where the beam pointing direction had no effect on the received SNR. Histograms of the velocity of the RX van in the LOS and NLOS runs are provided in Fig. \ref{fig:LOS_velocity} and Fig. \ref{fig:NLOS_velocity}. The average vehicular speed was 10.3 mph in LOS scenarios and 10.5 mph in NLOS scenarios.

\begin{table}[]
	\centering
			\caption{TX locations serving each LOS and NLOS run.}
	\label{tbl:Locations}
	\begin{tabular}{|c|c|}
		\hline
		\textbf{TX} & \textbf{Runs} \\ \specialrule{.2em}{.05em}{.05em} 
		TX 1 & NL1, NL2 \\ \hline
		TX 2 & L1, L2, NL3 \\ \hline
		TX 3 & L3, L4, NL4, NL5 \\ \hline
		TX 4 & L5, NL6 \\ \hline

	\end{tabular}
\end{table}

\section{Performance Impact of Beam Management} \label{sec:rx_design}
Beam management is the process of acquiring and maintaining beam pair links between a BS and user equipment (UE)\cite{Li_2020}. To determine the beam link directions with maximum SNR, the BS and UE sweep across different direction pairs. At each beam direction pair, the BS (or UE) transmits a known sounding signal, which is measured at the UE/BS to determine the channel SNR. Three types of beam sweeping procedures are discussed in 3GPP 5G NR standard \cite{3GPP.38.802}. In Procedure 1 (P1) beam sweeping during initial access, both the base station (BS) and the UE sweep through all possible beam directions to determine the optimal pointing direction for data transmission. In Procedure 2 (P2), beam sweeping occurs only at the BS. Once the link is established, Procedure 3 (P3) beam sweeping maintains the link, with beam sweeping at the UE only\cite{Li_2020}. We shall investigate P3 beam sweeping and its effect on the data communication rate based on the data collected from the outdoor vehicular measurements. 
\subsection{System Model}\label{sec:Model}

We shall now provide a system model to predict the average data throughput a vehicular UE would receive, based on the received SNR and the beam sweep period.

Let there be $ M $ beam directions at the vehicular UE, over which the UE must search for the best beam direction. Let $ T_s $ be the time required for the UE to switch from one beam direction to another and dwell in the new beam direction to measure the sounding signal. Assuming an exhaustive search across all beam directions, $ M \cdot T_s $ will be spent in beam alignment. If the beam sweep period is $ T $, a time period of $ T- M \cdot T_s$ is available for data transmission. The fraction of time available for data transmission is thus \cite{Destino_2017}
\begin{align}
\eta = 1-\dfrac{M\cdot T_s}{T}
\end{align}

With a longer beam sweep period, a greater fraction of time is available for data transmission. However, if beam sweeping is not done at a sufficient rate, the SNR at the UE will degrade as the vehicular environment changes due to mobility which then requires a change in the optimal beam pointing direction. 

The total data rate, $ R $, of a UE is given by
\begin{align}
R = \eta B \log_{2}\left(1+\text{SNR}\right),
\end{align}
where $ B $ is the RF bandwidth of the channel in Hertz, SNR is the signal-to-noise ratio at the UE (expressed as a linear ratio) \cite{Destino_2017}. We wish to obtain the beam sweep period $T$ that maximizes the UE data rates. 

\subsection{Optimal Beam Switching Speed}

\begin{figure}
	\centering
	\includegraphics[width=0.5\textwidth]{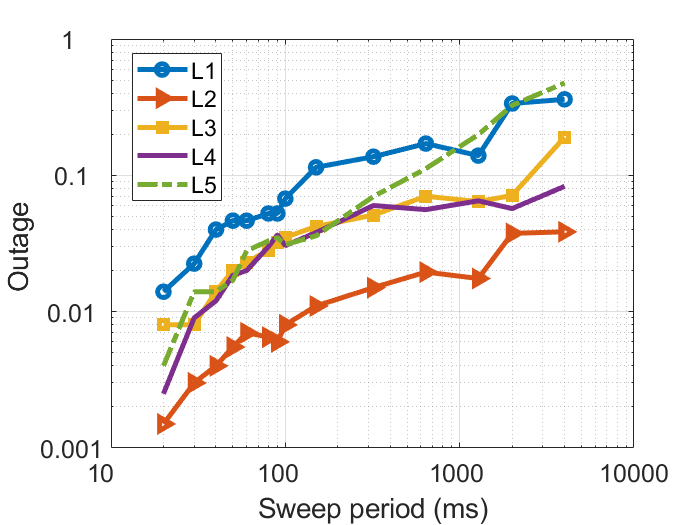}
	\caption{Outage likelihood vs beam sweep period in LOS.}
	\label{fig:LOS_outage} 	
\end{figure}

\begin{figure}
	\centering
	\includegraphics[width=0.5\textwidth]{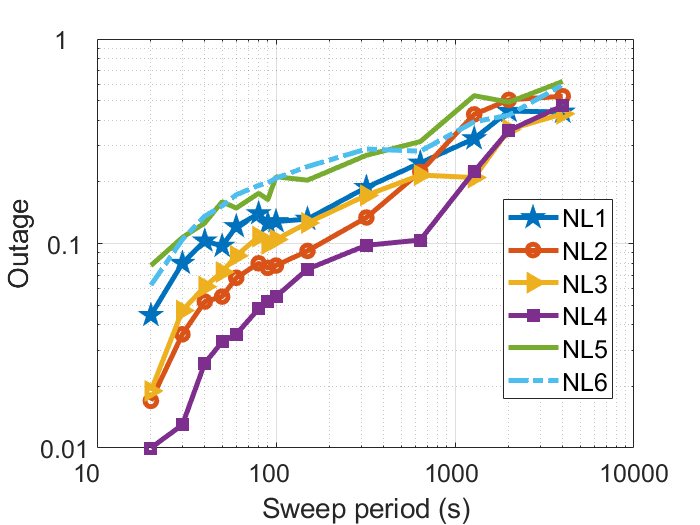}
	\caption{Outage likelihood vs beam sweep period in NLOS.}
	\label{fig:NLOS_outage} 	
\end{figure}

The modulation and coding scheme (MCS) choice at the RX is determined by the measured SNR. Although higher-order modulation schemes support higher data rates, a higher SNR is required to ensure low block error rates (BLER). If the duration between two channel measurements is too long, the SNR may drop to a level below the minimum SNR required to support the chosen MCS, thus leading to high BLER. Frequent channel measurements (i.e. a short beam sweep period) ensure that the RX correctly estimates the maximum MCS supported by the wireless channel. An outage event is defined as when the RX SNR falls 5 dB below the SNR measured during beam sweeping. As seen in Fig. \ref{fig:LOS_outage}, in LOS channels, if the interval between consecutive channel measurements in a particular beamforming direction, i.e. the beam sweep period, is greater than one second, an outage may occur more than 10$\%$ of the time in some scenarios (scenarios L1 and L5). The SNR in NLOS channels varies more rapidly as seen in Fig. \ref{fig:NLOS_outage}.


As discussed in Section \ref{sec:Model}, since the RX cannot transmit data while scanning the 360$ ^\circ $ azimuth and $ \pm 30^\circ $ elevation beam-space, there is a trade-off between scanning the beam-space for the beam pointing direction with maximum SNR and dwelling at a selected beam direction to transmit data. If the scanning rate is too high, the RX will spend a large percentage of time scanning the channel, without transmitting data, due to which the net throughput would drop. If the scanning rate is too low, the RX will miss beam directions with strong multipath. 

\begin{figure}
	\centering
	\includegraphics[width=0.5\textwidth]{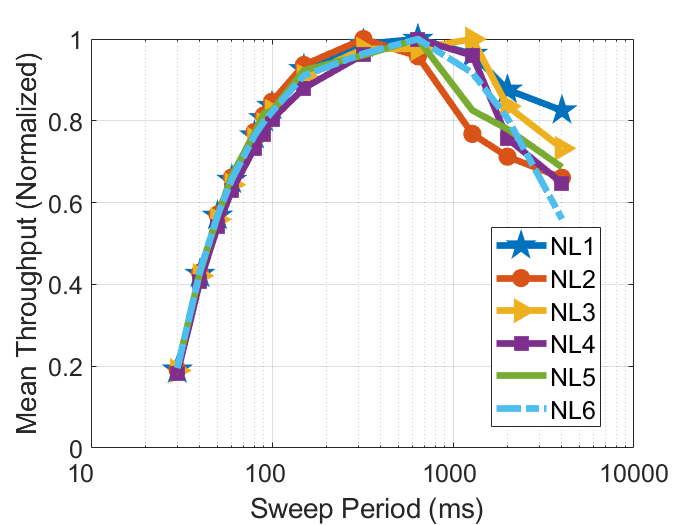}
	\caption{The normalized mean throughput for the NLOS measurements follows a concave trend with the beam sweep period. To maximize the mean throughput, a beam sweep period of 300 ms should be utilized.}
	\label{fig:optimal_switching} 	
\end{figure}

The impact of the beam sweep period on the mean throughput in the six NLOS runs is depicted in Fig. \ref{fig:optimal_switching}. The mean throughputs are normalized in order to isolate the effect of beam sweep period across runs with different SNR. For the six NLOS runs considered, the maximum data throughput was achieved with different beam sweep period. However, to maximize the mean throughput on average, a beam sweep period of 300 ms should be utilized. No appreciable effect of changing the beam sweep period was observed for the LOS runs which implies that the SNR of the LOS channels did not vary appreciably over the 10-20 second runs considered over a distance of 40 to 90 meters. 

The speed of the vehicle affects the optimal beam sweeping speed - vehicles moving at faster speeds must also sweep the channel at a faster rate. The velocity at which the vehicle was moving during the measurement runs, obtained from the GPS of the vehicle, is provided in Fig. \ref{fig:LOS_velocity} and Fig. \ref{fig:NLOS_velocity}.

%
\subsection{Impact of Number of RX Chains}
Directional antennas are used at mmWave frequencies to combat the larger path loss in the first meter of propagation \cite{Rappaport_2015,Rappaport_2017}. Consequentially, in order to receive signals from all directions, a typical vehicle with mmWave capabilities will have multiple RX panes. ROACH has four RX panes, one pointing to the front, the back, and to the sides of the vehicle. Data from each RX pane is synchronously collected via independent RX chains. The number of RX chains impacts the cost of deployment and hence a common RX chain may be used for multiple  RX panes in future vehicles. As a result, the RX must sequentially sweep over beams in each RX pane while searching for the optimal beam direction, leading to longer sweep times and a drop in overall throughput. Since each of the four RX panes had 50 potential beam directions and the beams were switched every 125 $ \mu $s, the fraction of time available for data transmission ($ \eta $) is given by
\begin{align}
	 \eta = 1- \dfrac{4\times 6.25 }{n_{RX chains}\times T}, 
\end{align} 
where $  n_{RX chains} $ is the number of RX chains (1, 2, or 4) and $ T $ is the beam sweep period in milliseconds. In addition to the lower beam sweeping overhead, $ n_{RX} $ RX chains may simultaneously monitor $ n_{RX} $ beam directions, the additional RX chains offer more resilience to blockages.

\begin{figure}
	\centering
	\includegraphics[width=0.5\textwidth]{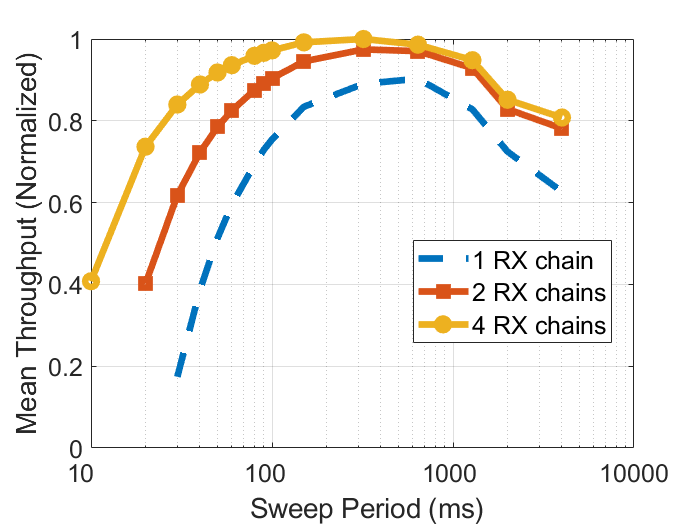}
	\caption{Mean throughput averaged over all NLOS runs. Increasing the number of RX chains leads to greater data transfer since the beam sweeping time decreases.}
	\label{fig:rx_chain_results} 	
\end{figure}

The performance of one, two, and four RX chains, averaged across the six NLOS runs, is shown in Fig. \ref{fig:rx_chain_results}. Utilizing one RX chain instead of four led to 10\% drop in the mean normalized throughput. However, four RX chains were more tolerant to varying the beam sweep period. A throughput  greater than 95\% of the maximum throughput was achieved with four RX chains for beam sweep periods ranging from 100 ms to 1000 ms.

\section{Conclusions and Future Work}\label{sec:conclusion}
MmWave frequencies have the potential to offer high data-rates for real-time vehicular sensor data sharing, however, narrow antenna beamwidths are required to compensate for the additional path loss in the first meter of propagation. To prevent link failure caused by dynamic blockage of the narrow beams, robust beam management algorithms are required. In this paper, based on real-world data gathered in an outdoor urban light commercial area, a beam sweeping rate of 300 ms was found to be optimal for mmWave V2X communications at an average vehicular speed of 10.5 mph. The wideband ROACH TX sounded the channel and the ROACH RX, which consists of a system of four RX phased arrays, scanned the channel 360$^\circ$ in the azimuth and 30$^\circ$ above and below the boresight. Six NLOS runs were used to determine the optimal sweeping rate, with each run being 10-20 seconds long, over a distance of 40 to 90 meters. Future V2I communication standards may leverage the optimal beam sweeping rate determined in this work for the design of efficient beam management algorithms.

Vehicles equipped with mmWave communication systems will have multiple RX panes in order to receive signals from all directions. The cost of mmWave vehicular deployment can be reduced by switching between multiple RX phased arrays with a single RX chain. Reducing the number of RX chains from four to one led to only a 10\% reduction in the mean throughput, normalized across all measurement runs, however there was vast robustness in throughput over an order of magnitude of beam sweep rate (100 ms to 1000 ms), showing the advantage of 4 RX chains.

The analysis in this paper assumed a constant beam sweeping rate. The vehicular surroundings influence the optimal beam sweeping period, with faster rates required in cluttered environments, or when the orientation of the vehicle is changing rapidly. The effect of aperiodic beam sweeping is left for future work.

\bibliographystyle{IEEEtran}
\bibliography{references}{}

\begin{thebibliography}{10}
\providecommand{\url}[1]{#1}
\csname url@samestyle\endcsname
\providecommand{\newblock}{\relax}
\providecommand{\bibinfo}[2]{#2}
\providecommand{\BIBentrySTDinterwordspacing}{\spaceskip=0pt\relax}
\providecommand{\BIBentryALTinterwordstretchfactor}{4}
\providecommand{\BIBentryALTinterwordspacing}{\spaceskip=\fontdimen2\font plus
\BIBentryALTinterwordstretchfactor\fontdimen3\font minus
  \fontdimen4\font\relax}
\providecommand{\BIBforeignlanguage}[2]{{%
\expandafter\ifx\csname l@#1\endcsname\relax
\typeout{** WARNING: IEEEtran.bst: No hyphenation pattern has been}%
\typeout{** loaded for the language `#1'. Using the pattern for}%
\typeout{** the default language instead.}%
\else
\language=\csname l@#1\endcsname
\fi
#2}}
\providecommand{\BIBdecl}{\relax}
\BIBdecl

\bibitem{Choi_2016}
J.~{Choi} \emph{et~al.}, ``Millimeter-wave vehicular communication to support
  massive automotive sensing,'' \emph{IEEE Communications Magazine}, vol.~54,
  no.~12, pp. 160--167, Dec. 2016.

\bibitem{Shimizu_2018}
T.~{Shimizu}, V.~{Va}, G.~{Bansal}, and R.~W. {Heath}, ``{Millimeter Wave V2X
  Communications: Use Cases and Design Considerations of Beam Management},'' in
  \emph{2018 Asia-Pacific Microwave Conference (APMC)}, Nov. 2018, pp.
  183--185.

\bibitem{USDoT}
{US Department of Transportation}. {Vehicle-to-vehicle Communication
  technology}.
  \url{https://www.nhtsa.gov/sites/nhtsa.dot.gov/files/documents/v2v_fact_sheet_101414_v2a.pdf}.
  Accessed: 2020-10-11.

\bibitem{Perfecto_2017}
C.~{Perfecto}, J.~{Del Ser}, M.~{Bennis}, and M.~N. {Bilbao}, ``{Beyond
  WYSIWYG: Sharing contextual sensing data through mmWave V2V
  communications},'' in \emph{2017 EuCNC}, June 2017, pp. 1--6.

\bibitem{fcc_cv2x}
\BIBentryALTinterwordspacing
``{Use of the 5.850-5.925 GHz Band, ET Docket No. 19-308}.'' [Online].
  Available: \url{https://docs.fcc.gov/public/attachments/FCC-19-129A2.pdf}
\BIBentrySTDinterwordspacing

\bibitem{Naik_2019}
G.~{Naik}, B.~{Choudhury}, and J.~{Park}, ``{IEEE 802.11bd 5G NR V2X: Evolution
  of Radio Access Technologies for V2X Communications},'' \emph{IEEE Access},
  vol.~7, pp. 70\,169--70\,184, May 2019.

\bibitem{Rappaport_2013b}
T.~S. Rappaport \emph{et~al.}, ``{Millimeter Wave Mobile Communications for
  {5G} Cellular: It Will Work!}'' \emph{IEEE Access}, vol.~1, pp. 335--349, May
  2013.

\bibitem{Rappaport_2015}
T.~S. Rappaport, R.~W. Heath~Jr, R.~C. Daniels, and J.~N. Murdock,
  \emph{Millimeter wave wireless communications}.\hskip 1em plus 0.5em minus
  0.4em\relax Pearson Education, 2015.

\bibitem{Rappaport_2017}
T.~S. {Rappaport}, Y.~{Xing}, G.~R. {MacCartney}, A.~F. {Molisch},
  E.~{Mellios}, and J.~{Zhang}, ``Overview of millimeter wave communications
  for fifth-generation (5g) wireless networks—with a focus on propagation
  models,'' \emph{IEEE Transactions on Antennas and Propagation}, vol.~65,
  no.~12, pp. 6213--6230, Aug. 2017.

\bibitem{Chopra_2020}
A.~Chopra \emph{et~al.}, ``Real-time millimeter wave omnidirectional channel
  sounder using phased array antennas,'' in \emph{2020 IEEE GLOBECOM}, Dec.
  2020, pp. 1--6.

\bibitem{MacCartney_2017}
G.~R. {MacCartney}, T.~S. {Rappaport}, and S.~{Rangan}, ``Rapid fading due to
  human blockage in pedestrian crowds at 5g millimeter-wave frequencies,'' in
  \emph{GLOBECOM 2017 - 2017 IEEE Global Communications Conference}, 2017, pp.
  1--7.

\bibitem{Park_2017}
J.~{Park} \emph{et~al.}, ``{Millimeter Wave Vehicular Blockage Characteristics
  Based on 28 GHz Measurements},'' in \emph{2017 IEEE 86th Vehicular Technology
  Conference (VTC-Fall)}, Sept. 2017, pp. 1--5.

\bibitem{Rappaport_2017_b}
T.~S. {Rappaport}, G.~R. {MacCartney}, S.~{Sun}, H.~{Yan}, and S.~{Deng},
  ``{Small-Scale, Local Area, and Transitional Millimeter Wave Propagation for
  5G Communications},'' \emph{IEEE Transactions on Antennas and Propagation},
  vol.~65, no.~12, pp. 6474--6490, Aug. 2017.

\bibitem{Akoum_2018}
S.~{Akoum}, A.~{Thornburg}, X.~{Wang}, and A.~{Ghosh}, ``Robust beam management
  for mobility in mmwave systems,'' in \emph{2018 52nd Asilomar Conf. on
  Signals, Systems, and Computers}, Oct. 2018, pp. 1269--1273.

\bibitem{Sun_2013}
{Shu Sun} and T.~S. {Rappaport}, ``{Multi-beam antenna combining for 28 GHz
  cellular link improvement in urban environments},'' in \emph{2013 IEEE
  GLOBECOM}, Dec. 2013, pp. 3754--3759.

\bibitem{Guo_2019}
Y.~{Guo}, Z.~{Wang}, M.~{Li}, and Q.~{Liu}, ``{Machine Learning Based mmWave
  Channel Tracking in Vehicular Scenario},'' in \emph{2019 IEEE ICC Workshops},
  May 2019, pp. 1--6.

\bibitem{Jayaprakasam_2017}
S.~{Jayaprakasam}, X.~{Ma}, J.~W. {Choi}, and S.~{Kim}, ``{Robust Beam-Tracking
  for mmWave Mobile Communications},'' \emph{IEEE Communications Letters},
  vol.~21, no.~12, pp. 2654--2657, Dec. 2017.

\bibitem{Fazliu_2020}
Z.~L. {Fazliu}, F.~{Malandrino}, C.~F. {Chiasserini}, and A.~{Nordio},
  ``{MmWave Beam Management in Urban Vehicular Networks},'' \emph{IEEE Systems
  Journal}, pp. 1--12, June 2020.

\bibitem{Va_2017}
V.~{Va}, J.~{Choi}, and R.~W. {Heath}, ``The impact of beamwidth on temporal
  channel variation in vehicular channels and its implications,'' \emph{IEEE
  Trans. on Vehicular Technology}, vol.~66, no.~6, pp. 5014--5029, 2017.

\bibitem{Hur_2018}
S.~{Hur} \emph{et~al.}, ``Feasibility of mobility for millimeter-wave systems
  based on channel measurements,'' \emph{IEEE Communications Magazine},
  vol.~56, no.~7, pp. 56--63, 2018.

\bibitem{Sun_2014}
S.~{Sun}, T.~S. {Rappaport}, R.~W. {Heath}, A.~{Nix}, and S.~{Rangan}, ``Mimo
  for millimeter-wave wireless communications: beamforming, spatial
  multiplexing, or both?'' \emph{IEEE Commun. Mag.}, vol.~52, no.~12, pp.
  110--121, Dec. 2014.

\bibitem{Wang_2020}
S.~Wang, J.~Huang, and X.~Zhang, ``Demystifying millimeter-wave v2x: Towards
  robust and efficient directional connectivity under high mobility,'' in
  \emph{ACM Mobicom}, Sept. 2020, pp. 1--14.

\bibitem{hua_2014}
M.~Hua, M.~Wang, K.~W. Yang, and K.~J. Zou, ``Analysis of the frequency offset
  effect on zadoff--chu sequence timing performance,'' \emph{IEEE Trans. on
  Communications}, vol.~62, no.~11, pp. 4024--4039, 2014.

\bibitem{STA}
``{Federal Communications Commission Experimental Radio Station Permit and
  License},'' \url{https://apps.fcc.gov/els/GetAtt.html?id=235534&x=.},
  accessed: 2020-05-23.

\bibitem{Li_2020}
Y.~R. {Li}, B.~{Gao}, X.~{Zhang}, and K.~{Huang}, ``{Beam Management in
  Millimeter-Wave Communications for 5G and Beyond},'' \emph{IEEE Access},
  vol.~8, pp. 13\,282--13\,293, 2020.

\bibitem{3GPP.38.802}
3GPP, ``{TSG RAN;Study on New Radio Access Technology Physical Layer Aspects
  (Release 14)},'' TR 38.802 V14.2.0, Sept. 2017.

\bibitem{Destino_2017}
G.~{Destino} and H.~{Wymeersch}, ``On the trade-off between positioning and
  data rate for mm-wave communication,'' in \emph{2017 IEEE ICC Workshops}, May
  2017, pp. 797--802.

\end{thebibliography}

\end{document}